\input epsf
\documentstyle[prd,floats,aps]{revtex}

\begin{document}
\title{\bf Colliding black holes with linearized gravity}

\author{Jorge Pullin\\  {\em Center for Gravitational Physics and Geometry}\\
{\em Department of Physics, 104 Davey Lab,}\\ {\em The Pennsylvania State
University,}\\
{\em University Park, PA 16802}}

\maketitle
\begin{abstract}

We give a brief summary of results and ongoing research in the
application of linearized theory to the study of black hole collisions
in the limit in which the holes start close to each other.  This
approximation can be a valuable tool for comparison and code-checking
of full numerical relativity computations. The approximation works
quite well for the head-on case and this is motivation to pursue its
use in other more interesting contexts.  We summarize current efforts
towards establishing the domain of validity of the approximation and
its use in generation and evolution of initial data for more
interesting physical cases.

\end{abstract}

\vspace{-8.0cm}
\begin{flushright}
\baselineskip=15pt
CGPG-95/8-2  \\
gr-qc/9508026\\
\end{flushright}
\vspace{7.0cm}
\twocolumn

\section{Introduction}

The intention of this talk is to summarize the application of linearized
gravity, in the specific form of the theory of black hole perturbations,
to the study of the collision of black holes. Most of the
results are already present in the literature, and the rest of the work
is still in progress so I present here only a brief survey.

The motivation for studying black hole collisions is quite clear. In
the next few decades gravitational wave detectors will come online
that will require ``templates'' of possible waveforms from different
sources.  The collision of black holes is one of the main candidates
for observable sources of gravitational radiation. Although the
initial and advanced LIGO detectors will not quite have the frequency
range to detect the waves produced in the final moments of the most
common collisions, it is expected that future detectors will, and
knowing the waveform for the final moments can also lead to insights
into the waveforms emitted earlier on.

The presence of this strong motivation from the experimental side has
led to the formation of an alliance of numerical relativity groups
(the ``binary black hole grand challenge collaboration'') with the
goal of numerically simulating the collision of two black holes using
supercomputers. The degree of difficulty of this project is reflected
in the fact that several established numerical relativity groups have
decided to team efforts in order to tackle it.

Here we will like to offer a much more modest approach, which is based
on a simple idea: when a collision of two black holes starts with the
holes so close to each other that they are surrounded by a common
horizon, the problem looks from the point of view of an external
observer as a single distorted black hole. It can therefore be treated
with perturbation theory. Although one expects this approach to only
yield results in a small range of initial separation, it provides ---at
least for that range--- a benchmark against which one can calibrate
numerical codes of the fully numerical approach.  In reference
\cite{PrPu} an explicit calculation was carried out using this
idea. We took the initial data for the head-on collision of two black
holes given by the Misner \cite{Mi} solution and re-wrote it in such a
way that in the case that the two black holes are close to each other
it explicitly looks like ``Schwarzschild plus something small". We
took the ``something small" and evolved it using the equations of
linearized gravity (the Zerilli equation) and computed the radiated
energy. The results are shown in figure 1, where we plot the energy
radiated in the collision as a function of the initial separation and
compare with the results of the NCSA group \cite{NCSA} using a
numerical integration of the full Einstein equations. We see that the
close approximation works very well until the holes are no longer
surrounded by a common apparent horizon ($\mu_0= 1.3$) and works
within the correct order of magnitude up to when the holes are no
longer surrounded by an event horizon ($\mu_0=2.0$). Also shown is a
``far approximation'' based on a particle-membrane paradigm
\cite{collab}. Comparisons of waveforms have also been performed
\cite{collab} and they also show very good agreement between the
linearized theory and the full numerical simulations.

\begin{figure}
\vspace{-2.5cm}
\epsfxsize=200pt \epsfbox{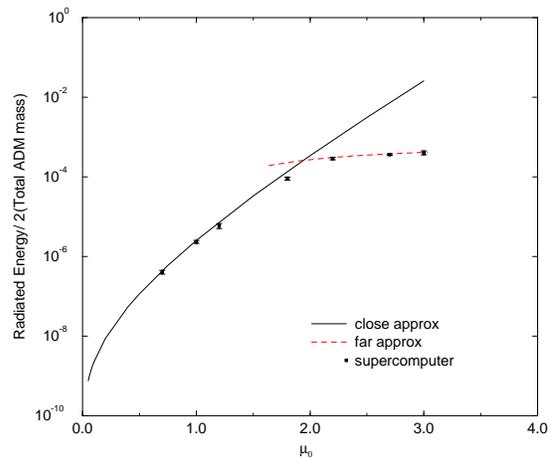}
\vspace{-0.5cm}
\caption{Comparison of results for the radiated
energy vs initial separation in a collision for the close
approximation, the fully numerical results and the far
approximation. Vertical scale is logarithmic. }
\end{figure}

All this shows that the use of linearized gravity in the close limit
can be a valuable aid to full numerical evolutions of the two black
hole problem. It is therefore quite tempting to apply the linearized
treatment to more interesting situations, specifically the
in-spiraling collision of two black holes with angular
momentum. There are two main obstacles to doing this computation and we
will detail them in the next two sections.

\section{Second order perturbations: giving the formalism
error bars}

Assuming initial data for a black hole collision is given, we can
rather easily evolve and compute energies in linearized theory. Why
therefore not do it for the in-spiraling collision? The main reason
is that for that case there are no numerical results with which to
compare and the linearized formalism does not have a measure of error
in it: it therefore has little predictive power. There is no
consistent way to say when the close approximation breaks down. In
fact, this example teaches us a valuable lesson about perturbation
theory: when is linearized perturbation valid? The obvious answer
``when perturbations are small'' is clearly naive. To begin with,
``small'' should be characterized in a coordinate invariant
way. Moreover, as this example shows, perturbations can be ``large''
and perturbation theory can still be valid: it just needs to happen
that the perturbations be large in regions of spacetime that do not
contribute in a significant way to the physics of interest. In the two
black hole example, such a region is the interior of the horizon and
regions close to it, in which perturbations mostly fall into the black
hole.

How is one to characterize when to trust the approximation? The answer is
simple: work out the second order perturbations, compute the physical
quantities of interest and use how much the first and second order
calculation differ as a measure of the accuracy of the first order
results. The advantage of this answer is that it is phrased in terms of
what one is exactly interested in: the physical quantities. In the case
of the collision of two black holes these are the radiated waveforms
and energies.

The formalism for second order perturbations of black holes has not been
worked out in the past. It can be studied in detail as we do in reference
\cite{zeri2}. Here I just sketch some of the outstanding points. It
turns out that all the information can be coded into a single variable,
exactly as in the first order perturbation case and that that variable
satisfies a ``Zerilli equation'',
\begin{equation}
-{\partial \psi^{(2)} \over \partial t} +
{\partial \psi^{(2)} \over \partial r_*} +V(r) \psi^{(2)} = S
\end{equation}
where $r_*=r+\log(r/2M-1)$ and the Zerilli function $\psi^{(2)}$ is a
coordinate invariant combination of the perturbed metric coefficients.
This equation is exactly the same as the one satisfied by the first
order perturbations (including the ``potential'' V(r), which can be seen
in reference \cite{PrPu}). However, there is an important difference:
the right-hand side is not zero but a ``source'' term S, which is listed
explicitly in reference \cite{zeri2} and which is a complicated function
quadratic in the first order perturbations and their derivatives. The
way in which we derived this equation is to compute a particular
combination of the Einstein equations, writing the perturbed metric in a
particular coordinate system, the so called ``Regge-Wheeler'' gauge.
This, in turn is a way of deriving the original Zerilli equation. The
expression we get for $\psi^{(2)}$ is therefore a representation in that
gauge of a gauge invariant quantity. The explicitly gauge invariant form
of $\psi^{(2)}$ can also be computed.

We therefore are in a position to evolve to second order the problem of
black hole collisions and therefore to endow the first order predictions
with ``error bars''. This will be crucial for the inspiralling case,
where numerical results are not expected for some time.

\section{Initial data in the close approximation}

In the head-on collision case we were lucky to have an exact solution to
the initial value problem that we could evolve. For the more realistic
cases there are no exact solutions available at present and it is
unlikely that they will be easily found in the future.
There is an immediate alternative at hand. There exist already well
tested numerical codes \cite{Cook} for solving the initial value
problem in general relativity in the context of black hole collisions.
One could simply take these initial data evaluated for the case in which
the black holes are close and ``read off'' from them the departures from
Schwarzschild to be evolved using the linearized theory. This is
certainly possible and has already been illustrated for
Brill-Lindquist-type initial data by Abrahams and Price \cite{AbPr}.

Apart from the possibility of using numerical initial data for
realistic collisions it is interesting to notice that one can, up to a
certain extent, solve the initial value problem analytically if one is
only interested in initial data for the close approximation. The idea
is simple: in the close approximation the initial data for a black
hole collision departs a small amount from the initial data for a
Schwarzschild spacetime for a single black hole with mass equal to the
sum of the masses of the colliding holes. Therefore one can develop an
approximation technique for the initial data starting from the initial
data of Schwarzschild and adding small corrections proportional to the
separation of the holes. We illustrate here only the zeroth order
results, details will be given in a forthcoming paper in collaboration
with John Baker.

The initial value problem of general relativity can be conveniently cast
in the conformal formalism \cite{Yo}. One is interested in solving the
momentum and Hamiltonian constraints
\begin{eqnarray}
\nabla^a (K_{ab} - g_{ab} K) &=& 0\\
{}^3R-K_{ab} K^{ab} + K^2 &=&0
\end{eqnarray}
where $g_{ab}$ is the spatial metric, $K_{ab}$ is the extrinsic
curvature and ${}^3R$ is the scalar curvature of the three metric. One
proposes a three metric that is conformally flat
$g_{ab} = \psi^4 \delta_{ab}$, with $\psi^4$ the conformal factor
and a decomposition of the extrinsic curvature
$\widehat{K}_{ab} = \psi^{-2} K_{ab}$.

The constraints become,
\begin{eqnarray}
\widehat{\nabla}^a \widehat{K}_{ab} &=& 0\\
\widehat{\nabla}^2 \psi &=& \psi^{-7} \widehat{K}_{ab} \widehat{K}^{ab}.
\end{eqnarray}
where $\widehat{\nabla}$ is a derivative with respect to the flat
spacetime. Since the momentum constraint is linear, one can propose as
a solution for it for the case of two black holes the sum of the
solutions for the case of individual holes\footnote{The particular
solution chosen depends on the boundary conditions imposed. This may
add other terms to the simple ones we list here for brevity, but they
all behave in a similar fashion with respect to the approximations we
will consider.} with momentum $P_a$,
\begin{equation}
\hat{K}_{ab} = {3 \over 2 r^2} \left[ P_{(a} n_{b)} -(\delta_{ab}
-n_a n_b)P^c n_c\right]
\end{equation}
where $n_{b}$ is a unit normal in the direction of $\vec{r}$ and all
vector fields are defined in the flat background spacetime.

One now can put this solution in the Hamiltonian constraint and one is
left with an elliptic, highly non-linear equation for $\psi$. This is
the equation that is usually solved numerically. There exist
situations, however, where one can make some progress
analytically. Consider the case in which the momenta of the holes is
small \cite{Yo}. In that case one can neglect the right-hand side of
the Hamiltonian constraint and one only needs to solve a vacuum
Laplace equation for $\psi$. The solution can therefore be very simply
found, the difficulty depending on the boundary conditions one chooses
for the problem (typically a ``symmetrized'' boundary condition is
imposed, which complicates calculation quite a bit in certain cases,
see \cite{Cook} for details).

\begin{figure}
\vspace{-2.5cm}
\epsfxsize=200pt \epsfbox{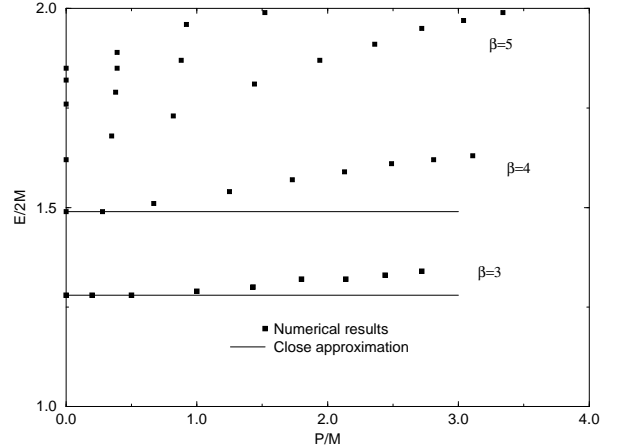}
\vspace{-0.5cm}
\caption{The ADM energy of
initial data for collisions of black holes of momenta
$P$. The dots are the full numerical results of Cook, for different
values of the initial separation $\beta$. We see that for small
separations, the energy is approximately independent of the holes
momenta, which coincides with the close approximation prediction,
depicted by the solid line.}
\label{cook}
\end{figure}

Another situation in which one can obtain an approximate solution is in
the ``close approximation''. In that case one has two black holes of
momenta equal and opposite $P^{(1)}_a=-P^{(2)}_a$, and since the black
holes are close, the unit normals appearing in the form for the
extrinsic curvature for each hole are approximately equal. That implies
that the extrinsic curvature for the problem is approximately zero (as
it should, since in the close limit the problem looks like a
Schwarzschild black hole at rest.) Therefore one can again neglect the
right-hand side of the Hamiltonian constraint and one is again left with
a Laplace equation. Let us compare this approximation with the full
numerical results. In order to do this we will compare the ADM energy of
initial data for a collision of two holes of momentum $P$. The ADM
energy in the conformal formalism is given by

\begin{equation}
E=-{1\over 2 \pi}\oint_\infty \nabla_i\psi \, d^2 S^i
\end{equation}
and we notice that it does not depend explicitly on the extrinsic
curvature (it does implicitly via the constraints). Therefore at the
approximation we are working, in which the constraints do not couple the
conformal factor and the extrinsic curvature, the energy is independent
of the extrinsic curvature and therefore independent of the momenta of
the holes. We compare this prediction with the full numerical results of
Cook in figure \ref{cook}.

An interesting aspect is that one can advance this approximation one
step further. One can input the extrinsic curvature and the conformal
factor found as a fixed ``source'' in the equation determining the
conformal factor and one can obtain a correction through the integration
of a Poisson equation. Comparison of this approximation with the
numerical data is currently in progress. Details are complicated by the
particular boundary conditions that are usually chosen in the numerical
computations.

It is evident that the ``close approximation'' can work in many other
cases, apart from the head-on, equal momenta holes we considered here.
The only changes will be that the solution one obtains in the ``close
limit'' rather than being a slice of Schwarzschild will be a slice of
Kerr or boosted Schwarzschild if the net result of the collision has
angular momentum or linear momentum.

\section{Summary}

We have seen that the use of the ``close approximation'' can be a
valuable aid to full numerical computations of the collision of two
black holes. With the introduction of a second order scheme we are now
in a position of offering reliable estimates of energies and waveforms
that we expect people working on the full numerical simulations will
find of use to calibrate codes and design strategies for better
integrating the Einstein equations in this problem of great current
physical interest.

\section*{Acknowledgments}

The work described here is in collaboration with Richard Price, John
Baker, Reinaldo Gleiser and Oscar Nicasio. I acknowledge support of NSF
through grants PHY94-06269 PHY93-96246, funds of the Pennsylvania State
University, its Office for Minority Faculty Development, and the Eberly
Family research fund.

\end{document}